# Broadband Acoustic Collimation and Focusing using Reduced Aberration Acoustic Luneburg Lens

Liuxian Zhao[1], Miao Yu[1,2,*]

[1]Institute for Systems Research, University of Maryland, College Park, MD, 20742, USA

[2]Department of Mechanical Engineering, University of Maryland, College Park, Maryland 20742, USA

*Author to whom correspondence should be addressed: mmyu@umd.edu

**ABSTRACT**

Acoustic Luneburg lens is a symmetric gradient-index lens with a refractive index decreasing radially from the centre to the outer surface. It can be used to manipulate acoustic wave propagation with collimation and focusing capabilities. However, previously studied acoustic Luneburg lens works only at audible frequency ranges from 1 kHz to 7 kHz, or at a single ultrasonic frequency of 40 kHz. In addition, the acoustic Luneburg lens at high frequency is not omnidirectional in the previous researches. Furthermore, there is no realization of simulation and experimental testing of 3D acoustic Luneburg lens until now. In this paper, a practical reduced aberration acoustic Luneburg lens (RAALL) are proposed for broadband and omnidirectional acoustic collimation and focusing with reduced aberrations. Ray tracing technique shows that RAALL can achieve a better acoustic focusing compared with traditional modified acoustic Luneburg lens. Following that, two models of RAALL are designed and fabricated through additive manufacturing technology: a 2D version and a 3D version. Collimation and focusing performances of the ultrasonic waves are theoretically, numerically and experimentally demonstrated for both 2D and 3D lenses, and their broadband and omnidirectional characteristics are verified.

## 1. Introduction

Luneburg lens is a spherical/circular lens with refractive index varies as a function of the radial distance from the centre of the lens, which was first proposed by Luneburg in the 1940's [1] and then extended by Gutman [2] and Morgan [3]. In recent years, the optical Luneburg lens has attracted widespread concerns in the application of electromagnetic wave transportation and communication system [4-8]. Most of previous researches have already been demonstrated by geometrical optics [9], which validate that electromagnetic plane wave propagates through a Luneburg lens can be focused at the axial point on the opposite side of the lens surface.

With the development of graded index acoustic metamaterials in recent years [10-19], Luneburg lens are getting considerable attentions not only from optical waves [20, 21], but also from elastic structures and acoustics [22-28]. For the elastic waves, Climente *et al*. fabricated a Luneburg lens for flexural waves focusing based on variations of plate's thickness [29]. Tol *et al*. explored a phononic crystal Luneburg lens using hexagonal unit cells with blind holes, which can achieve omnidirectional elastic wave focusing and energy harvesting [30]. Zhao *et al*. investigated a modified structural Luneburg lens, which can achieve elastic wave collimation and focusing with variable focusing length [31]. For acoustic waves, Kim *et al*. designed a two-dimensional acoustic Luneburg lens as sound reception system, which is made of 253 acrylic pipes of different radii [32]. The working frequency range of the lens is $f$ = 300 Hz - 900 Hz. In order to increase the working frequency,

Dong *et al*. designed a broadband two-dimensional flattened Luneburg lens for acoustic wave focusing, which works at the frequency range $f$ = 3 - 7 kHz [33]. Further, Xie *et al*. fabricated a 2D acoustic Luneburg lens for focusing, which works for $f$ = 40 kHz [34]. However, for this type of circular/spherical acoustic Luneburg lens, which only works at a single frequency. In addition, by exploring the equi-frequency contour (EFC) of the unit cell, it found that the designed structure is non-omnidirectional. Furthermore, there is neither numerical simulations nor experimental validations of the fabricated 3D structures.

In order to apply acoustic Luneburg lens for ultrasonic imaging, diagnosis, treatment and sonar system, broadband and omnidirectional characteristics of the lens are required. In this paper, we modify the traditional acoustic Luneburg lens (blue curve in Figure 1(b)), which is named modified acoustic Luneburg lens (black curve in Figure 1(c)). However, the modified acoustic Luneburg lens shows drastic ray aberrations (Figure 1(c)). Therefore, a novel governing equation for reduced aberration acoustic Luneburg lens (RAALL) is proposed to achieve reduced aberrations of acoustic collimation and focusing (red curve in Figure 1(d)). The governing equations of refractive index of the three lenses are presented in Figure 1(a). Full-wave simulations based on the finite element method are performed to validate the proposed design. Finally, a 2D and a 3D RAALLs are fabricated using additive manufacturing based on graded unit cells (3D truss) and tested in order to demonstrate the functionalities of the proposed lens at the ultrasonic frequency range $f$ = 20 - 40 kHz (Figure 1(e)).

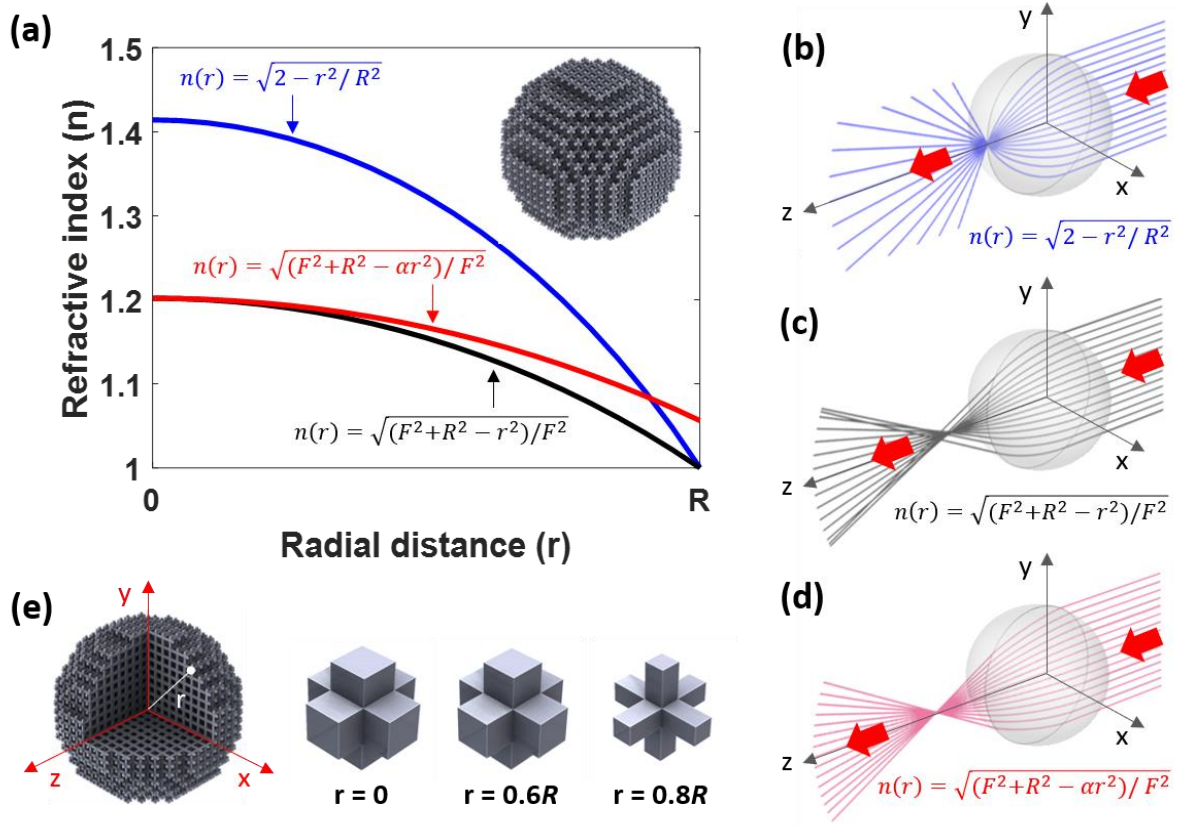

**Figure 1: Design principles of acoustic Luneburg lens. (a) The governing equations and their plots for traditional ALL, modified ALL, and RAALL. (b)-(d) Ray trajectories for traditional ALL, modified ALL, and RAALL, correspondingly. (e) Designed 3D RAALL with graded unit cells.**

## 2. Design Principles

The spatially varied refractive index of traditional acoustic Luneburg lens is defined as the following equation [33]:

$$n(r) = \sqrt{2 - (r/R)^2}, \quad 0 < r < R \quad , \tag{1}$$

The refractive index in equation (1) requires the refractive index to have a range between 1 and 1.42 (blue curve in Figure 1(a)). Kinds of unit cell structure are possible to achieve above refractive index range, here we use a unit cell of 3D truss [34, 35], which has three orthogonal beams, so that each cell is interconnected with its adjacent cells to form a self-supported lattice. In principle, a range of variable refractive index values can be obtained using this design via changing the unit cell filling ratio. In the following studies, $d$ is chosen to be 2 mm. The filling ratio of the unit cell can be changed via changing the factor $a_0$, and hence the refractive index can be changed correspondingly (Figure 1(e)). The refractive index of each unit cell is calculated based on a standard retrieval method [36].

The dispersion curve of the unit cell can be calculated using commercial software COMSOL. Figure 2(a) shows the first band of the dispersion relation along the $\Gamma X$ direction for the truss unit cell with various factor $a_0 = 0.1 - 0.7$. We can see that the dispersion curve is almost linear as $a_0$ is low in a broadband frequency range, which indicates the structure can be treated as homogenous as $a_0$ is low. Figure 2(b) presents the relation between the factor $a_0$ with the refractive index $n$ of the unit cell. We can see that in the low frequency region, the refractive index of the unit cell keeps constant since the dispersion relation is almost linear in this region. It means that the periodic structure can be considered as a homogeneous material in the low frequency region. In addition, the frequency band is much broader at low factor comparing with high factor. However, for the traditional acoustic Luneburg lens, which requires the maximum index goes up to 1.42. It can be seen from Figure 2(b) that the factor $a_0$ requires to be over 0.6, where the index is not linear with changing of frequency when $f >$ 30 kHz, as seen from Figure 2(a). That's the reason why in the literature [34], the acoustic Luneburg lens only works at a single frequency 40 kHz. In addition, EFC analysis is performed to analyse the omnidirectional characteristic of acoustic Luneburg lens. When the factor $a_0 = 0.6$, the unit cell is anisotropy when the frequency $f > 30$ kHz, as shown in Figure 2(c).

In order to extend the frequency range and improve the anisotropy limitation, one option is to reduce the maximum refractive index. A modified acoustic Luneburg lens is proposed based on the modified optical Luneburg lens (black curve in Figure 1(a)) [2]:

$$n(r) = \frac{\sqrt{F^2+R^2-r^2}}{F}, \qquad (2)$$

where $r$ is the radial distance from the central point, $R$ is the radius of the lens, and $F$ is the focusing length.

Based on the relation between the refractive index and focusing length, the maximum refractive index is 1.20 when $F = 1.5R$, which corresponds to a factor $a_0 = 0.4$ from Figure 2(b). It can be seen that the dispersion curve is almost linear when the frequency changing from 0 kHz – 40 kHz. Note that the available frequency range for collimation and focusing is limited by the structure of the lens as $4d < \lambda < R$ based on homogeneity condition of wave [37], where $\lambda$ is the wavelength, $d$ is the periodicity and $R$ is the radius of the lens. In this study, $d$ is 2 mm, and $R$ is 20 mm. Therefore, we choose the frequency range $f = 20 – 40$ kHz. In addition, it can be seen from Figure 2(d) that the EFC is almost circular when the frequency changing from 20 kHz to 40 kHz, which indicates that the lens is omnidirectional at this frequency range.

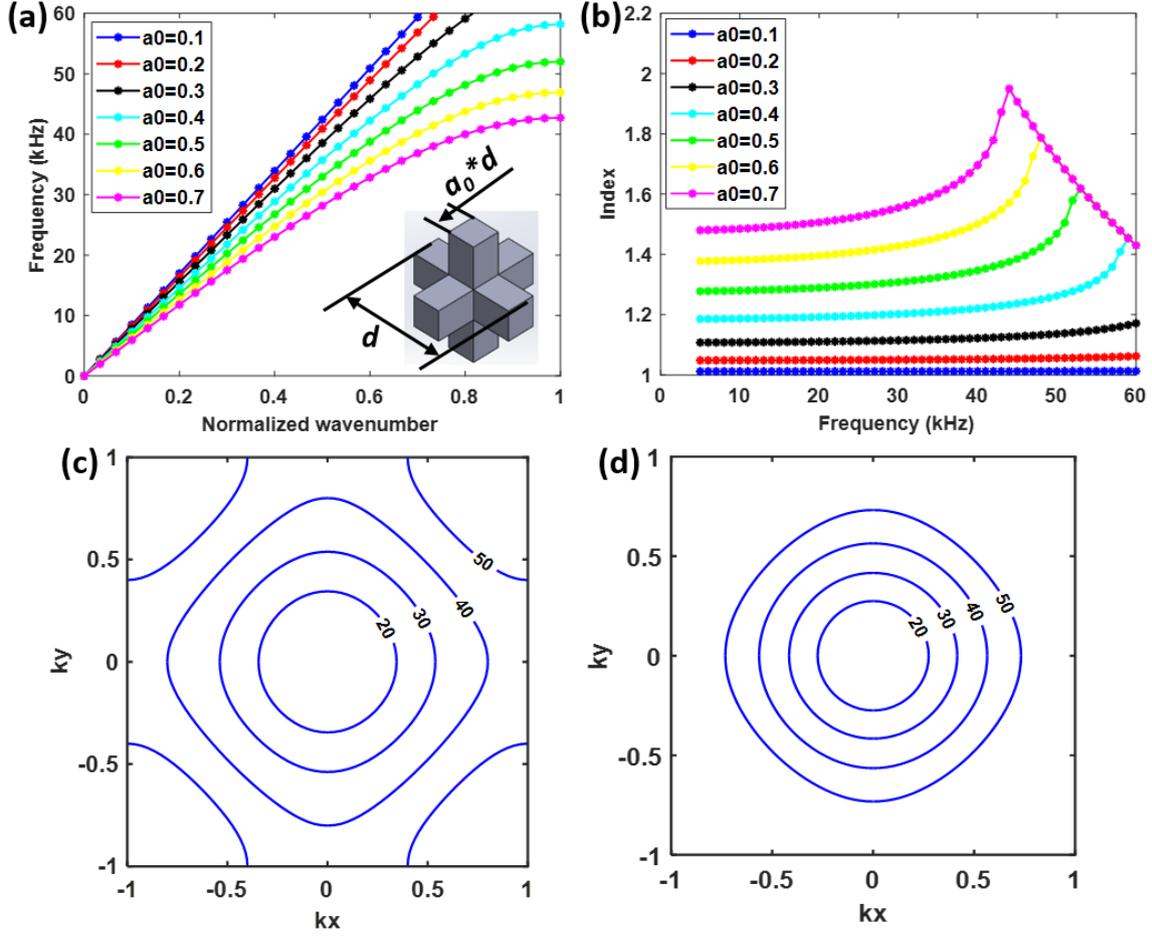

**Figure 2: Dispersion and homogeneity analysis of unit cell. (a) The effect of factor $a_0$ on the first band dispersion curve. The inset indicates an example of the truss unit cell. (b) The effect of factor $a_0$ on the refractive index. (c) Equi-frequency contour plot when $a_0$ = 0.6. (d) Equi-frequency contour plot when $a_0$ = 0.4.**

We calculate the ray trajectory of the modified ALL, as shown in Figure 3(a). However, it is clearly show that the modified ALL is capable of focusing an incident plane wave while with high ray aberration. In order to reduce the ray aberration of focusing, a reduced aberration acoustics Luneburg lens (RAALL) is defined in this study as (red curve in Figure 1(a)):

$$n(r) = \frac{\sqrt{F^2+R^2-\alpha r^2}}{F} \quad , \quad (3)$$

where $\alpha$ is a corrector.

We can see that an added corrector cannot change the maximum value of the refractive index, so the frequency range and omnidirectional characteristics are not changed. Now we define an error objective function to reduce the ray aberration based on:

$$Err = \sqrt{\frac{1}{N}\sum_{i=1}^{N}\{x_i(y=0)-u\}^2} \quad , \tag{4}$$

where $x(y = 0)$ indicates the intersection of the ray with $x$ axis (when $y = 0$), and $u$ is the mean value of $x(y = 0)$, and $N$ is the number of rays. COMSOL optimization is used to find the minimum value of the error objective function with varying corrector $\alpha$, which obtain a nearly zero aberration acoustic Luneburg lens when the $\alpha = 0.74$, with focusing location at $l = 1.87R$. We calculate the ray trajectory of the RAALL, which clearly shows an improved focusing with minimal ray aberration, as shown in Figure 3(b).

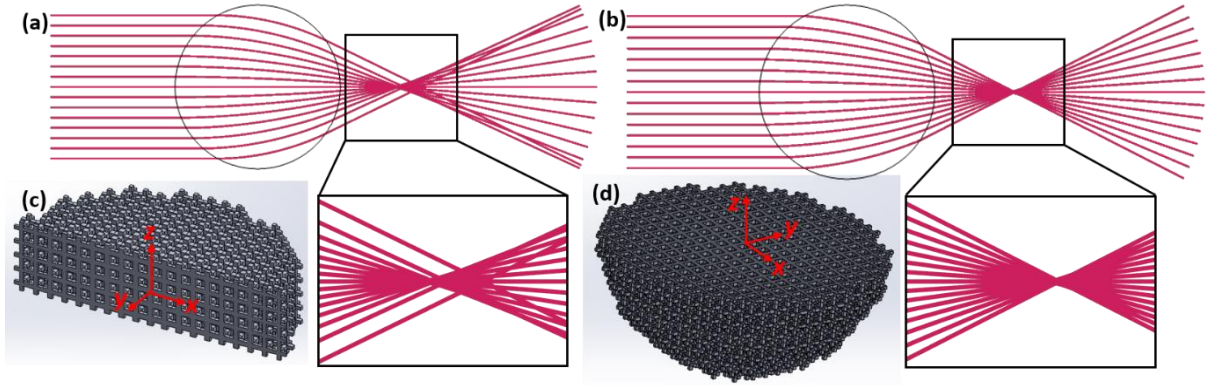

**Figure 3: Ray tracing comparison between modified acoustic Luneburg lens and RAALL, and cross-sections of 2D and 3D RAALLs. (a) Ray tracing for modified acoustic Luneburg lens. (b) Ray tracing for RAALL. (c) and (d) are the 2D version and 3D version RAALLs, correspondingly.**

In the following, we design and fabricate two RAALLs: a 2D version and a 3D version which work in broadband ultrasonic frequency range $f = 20$ kHz - 40 kHz for both collimation and focusing. The designs are based on a series of 3D truss unit cells that can be stacked layer-by-layer to form a stable lattice (5 layers for 2D version, and 19 layers for 3D version), as shown in Figure 3(c) and (d). Note that the 2D distribution of the refractive index is obtained via extending the index (Equation 3) in the thickness direction. The 3D distribution of refractive index is obtained by rotating the index (Equation 3) along its central axis ($x$ axis).

## 3. Numerical Simulations

In this study, we explore the acoustic wave collimation and focusing capabilities of the RAALL. Full 3D wave simulations are conducted using commercial software COMSOL for both 2D and 3D lenses. Radiation boundary conditions are applied on the outer boundaries to assume infinite air spaces. For acoustic collimation analysis, a point source located at $(x, y, z) = (-l, 0, 0)$ is used for excitation. The simulation results at three different frequencies $f$ = 20 kHz, 30 kHz, 40 kHz are shown in Figure 4(a)-(c) for 2D lens. The simulation results reveal the excellent capability of the RAALL to convert the point feeding source into plane wave forms. In order to analyse the omnidirectional characteristic of RAALL, a rotation of the RAALL with 45º along $z$ axis and conduct the full simulations. Similarly, the simulation results at the same three different frequencies and shown in Figure 3(d)-(f). These results show a trivial difference with the non-rotated lens, which validate that RAALL is an omnidirectional lens for acoustic collimation.

For acoustic focusing analysis, a plane wave front is used for excitation. The simulation results at three different frequencies $f$ = 20 kHz, 30 kHz, 40 kHz are shown in Figure 4(g)-(i) for the 2D lens. The simulation results reveal the excellent capability of the RAALL to transform the plane wave front to be focused in a small area. Similarly, in order to analyse the omnidirectional characteristic of RAALL for acoustic focusing, a rotated RAALL with 45º along *z* axis reveals the same focusing performance as the non-rotated lens (Figure 4(j)-(l)).

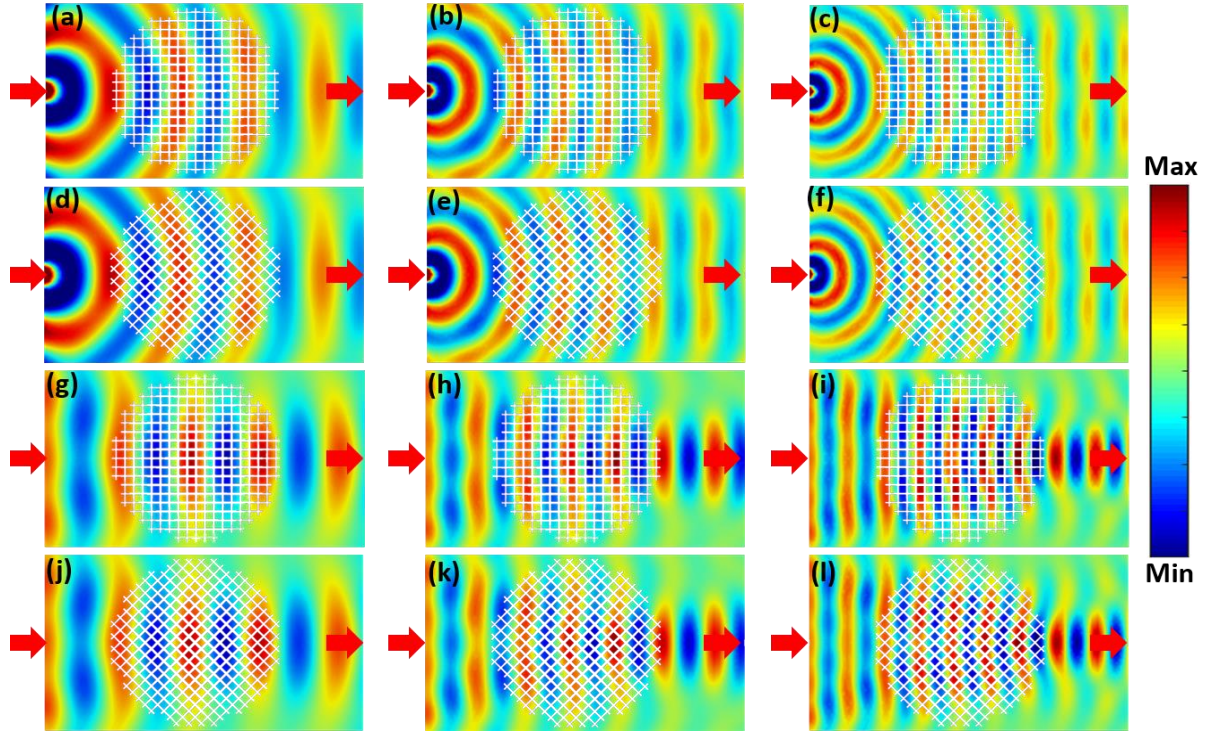

**Figure 4: Numerical simulation results for acoustic collimation and focusing using 2D lens. (a) – (c) are the waveform distributions at the frequencies *f* = 20 kHz, 30 kHz, 40 kHz correspondingly for acoustic collimation. (d) – (f) are the waveform distributions of the lens with a rotation of 45º along the *z* axis at the frequencies *f* = 20 kHz, 30 kHz, 40 kHz correspondingly for acoustic collimation. (g) – (i) are the waveform distributions at the frequencies *f* = 20 kHz, 30 kHz, 40 kHz correspondingly for acoustic focusing. (j) – (l) are the waveform distributions of the lens with a rotation of 45º along the *z* axis at the frequencies *f* = 20 kHz, 30 kHz, 40 kHz correspondingly for acoustic focusing.**

Furthermore, acoustic collimation and focusing analysis of the 3D lens are performed and shown in Figure 5. For acoustic collimation analysis, a point source located at (*x*, *y*, *z*) = (-*l*, 0, 0) is used for excitation. The simulation results at four different frequencies *f* = 20 kHz, 30 kHz, and 40 kHz are shown in Figure 5(a) - (c). The simulation results reveal the excellent capability of the 3D RAALL to convert the point feeding source into plane wave forms in a cylindrical area. For acoustic focusing analysis, a circular plane wave front is used for excitation. The simulation results at four different frequencies $f$ = 20 kHz, 30 kHz, and 40 kHz are shown in Figure 5(d)-(f). The simulation results reveal the excellent capability of the RAALL to transform the plane wave front to be focused in a small area.

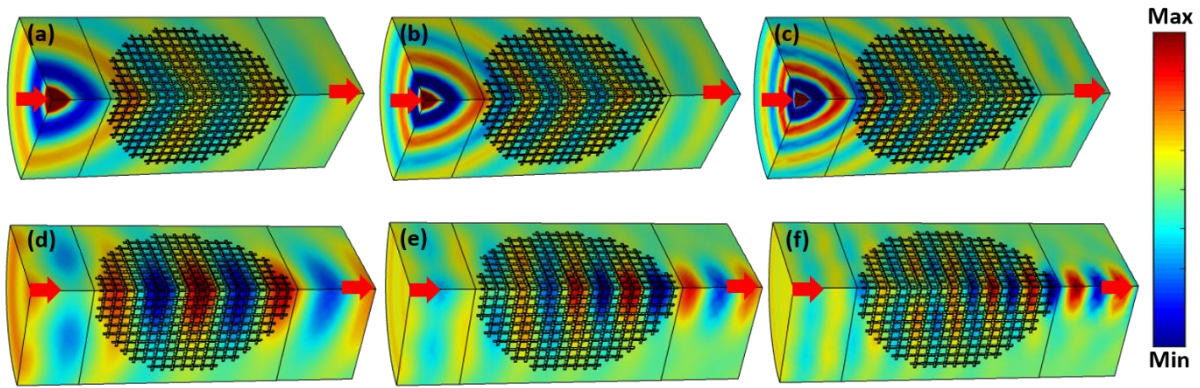

**Figure 5: Numerical simulation results for acoustic collimation and focusing using 3D lens. (a) – (c) are the waveform distributions of acoustic collimation at the frequency of $f$ = 20 kHz, 30 kHz, and 40 kHz correspondingly. (d) - (f) are the waveform distributions of acoustic focusing at the frequency of $f$ = 20 kHz, 30 kHz, and 40 kHz correspondingly.**

## 4. Experimental Verifications

To avoid inconsistency with the numerical simulations, in the experiments, the 2D and 3D RAALLs are fabricated with the same dimensions as the models in the simulations, as shown in Figure 6(a) and (b). Additive manufacturing technique is applied using Stratasys Objet500 Connex3 3D printer to fabricate the 2D and 3D lenses, which can offer a maximum resolution of 16 µm. Figure 6(c) and (d) show schematic and photo of the experimental setup for acoustic testing, which is performed in an echoless chamber. For acoustic collimation, a compact speaker (MA40S4S from Murata Manufacturing Co., Ltd.) located at $(x, y, z)$ = $(-l, 0, 0)$ is used to generate a point source for excitation. For acoustic focusing, a compact speaker located at $(x, y, z) = (-8R, 0, 0)$ is used to generate an approximate plane wave front for excitation, which is based on that a far field point source is approximate to a plane wave front. A fiber optic probe is used to measure the acoustic field distribution. A reference microphone (Type 4138 from Bruel & Kjaer) is used to collect the phase distribution. A three-dimensional stage is used to measure the acoustic field in three dimensional. A pulse signal with different frequencies is used as excitation source to the speaker with input voltage amplitude $V_p = 5$ V.

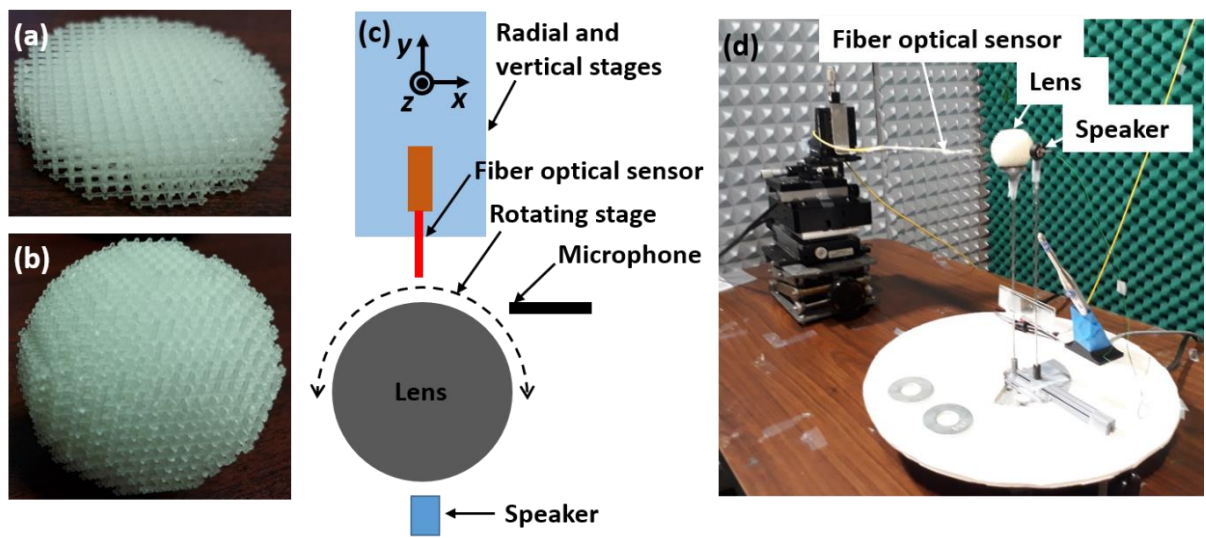

**Figure 6: Experimental setup for measurement of the full-field wave propagations in the scan area. (a) and (b) Photos of the 2D and 3D RAALLs, respectively. (c) A schematic of experimental setup, and (d) a photo of the overall experimental setup.**

We first test the 2D lens, for acoustic collimation, we measure both the acoustic amplitude and waveform of the pressure distributions with the presence of RAALL at the frequency of 40 kHz. It can be seen that the waveforms become a plane wave front with the presence of RAALL (Figure7(a)), which confirms that the RAALL has the capability of acoustic collimation. In order to investigate the broadband and omnidirectional characteristics of RAALL for acoustic collimation using 2D lens. A rotation of the lens with 45º is conducted at frequencies $f$ = 30 kHz and 40 kHz (Figure 7(b) and (c)). These results show a trivial difference with the non-rotated lens, which validate that RAALL is a broadband and omnidirectional lens for acoustic collimation.

In order to explore the acoustic focusing capability of RAALL using 2D lens. A far field source is used to generate an approximate plane wave front (the speaker located at ($x$, $y$, $z$) = (-8$R$, 0, 0)). The measurement results are shown in Figure 7(d)-(f). Figure 7(d) presents the acoustic waveform distribution at the frequency of 40 kHz without rotation of the lens. Figure 8(e) – (f) present the acoustic waveform distribution at the frequency of 30 kHz and 40 kHz with a rotation of 45º of the lens correspondingly. These results demonstrate that the RAALL is able to focus acoustic wave into a small area with broadband and omnidirectional characteristics.

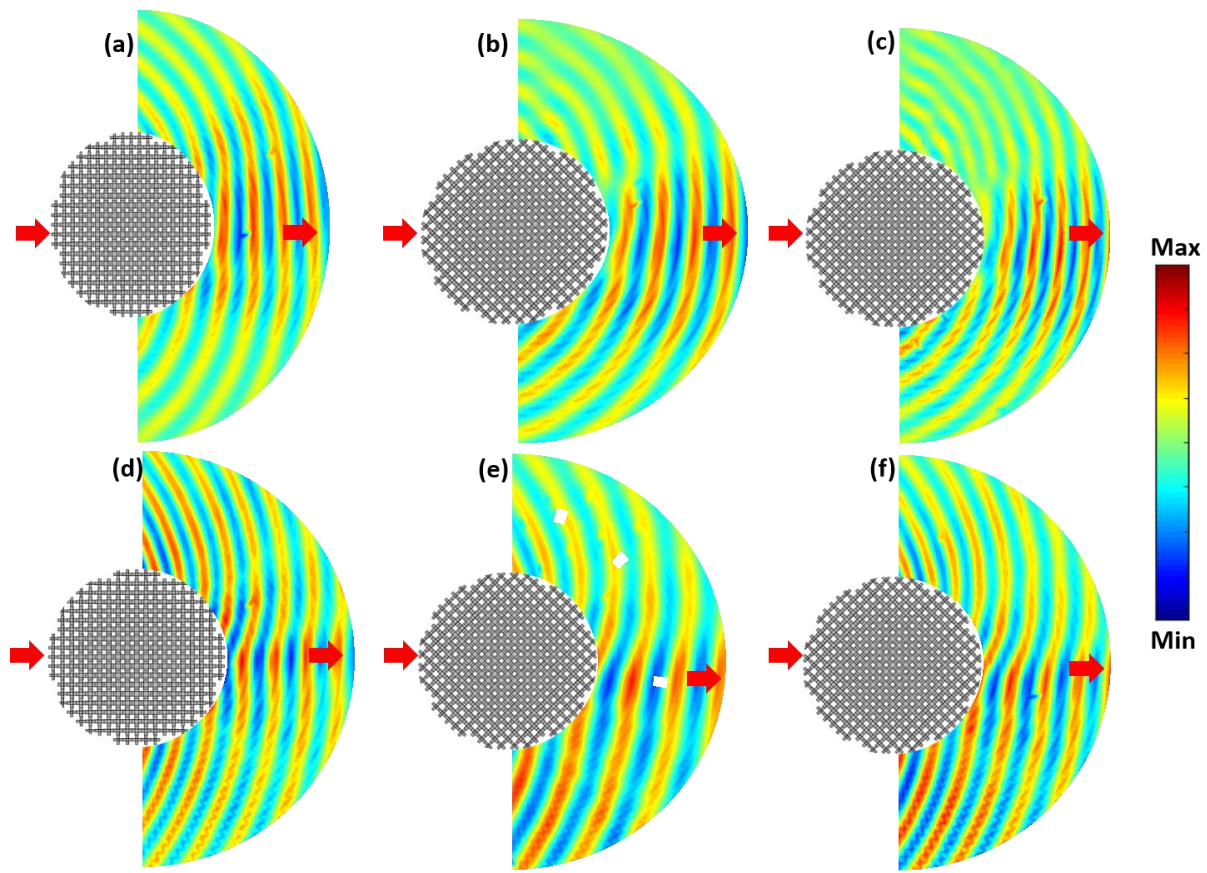

**Figure 7: Experimental results for acoustic collimation and acoustic focusing using 2D lens. (a) acoustic collimation at 40 kHz without rotation, (b) acoustic collimation at 30 kHz with a rotation of 45º, (c) acoustic collimation at 40 kHz with a rotation of 45º; (d) acoustic focusing at 40 kHz without rotation, (e) acoustic focusing at 30 kHz with a rotation of 45º, (f) acoustic focusing at 40 kHz with a rotation of 45º**

Similarly, in order to explore the acoustic collimation and focusing capabilities of RAALL using 3D lens. Both near field (the speaker located at (x, y, z) = (-$l$, 0, 0)) and far field (the speaker located at (x, y, z) = (-8$R$, 0, 0)) acoustic sources are used to generate a spherical wave front and an approximate plane wave front. The measurement data are collected from three surfaces located at $z$ = -2 mm, 0 mm and 2 mm. The experimental results are shown in Figure 8 present the acoustic wave front distribution at the frequency of $f$ = 40 kHz. Figure 8(a)-(c) show a clear collimated wave front after point source passes through 3D lens, and Figure 8(d)-(f) show a clear focused wave after plane wave passes through 3D lens. In addition, for the 3D lens design, it shows that the acoustic collimation and focusing not only works at the central surface ($z$ = 0 mm), but also works in the entire domain, such as $z$ = -2 mm and 2 mm. The experimental results keep good consistent with the numerical simulations and prove that both acoustic collimation and acoustic focusing can be obtained using the 3D RAALL lens.

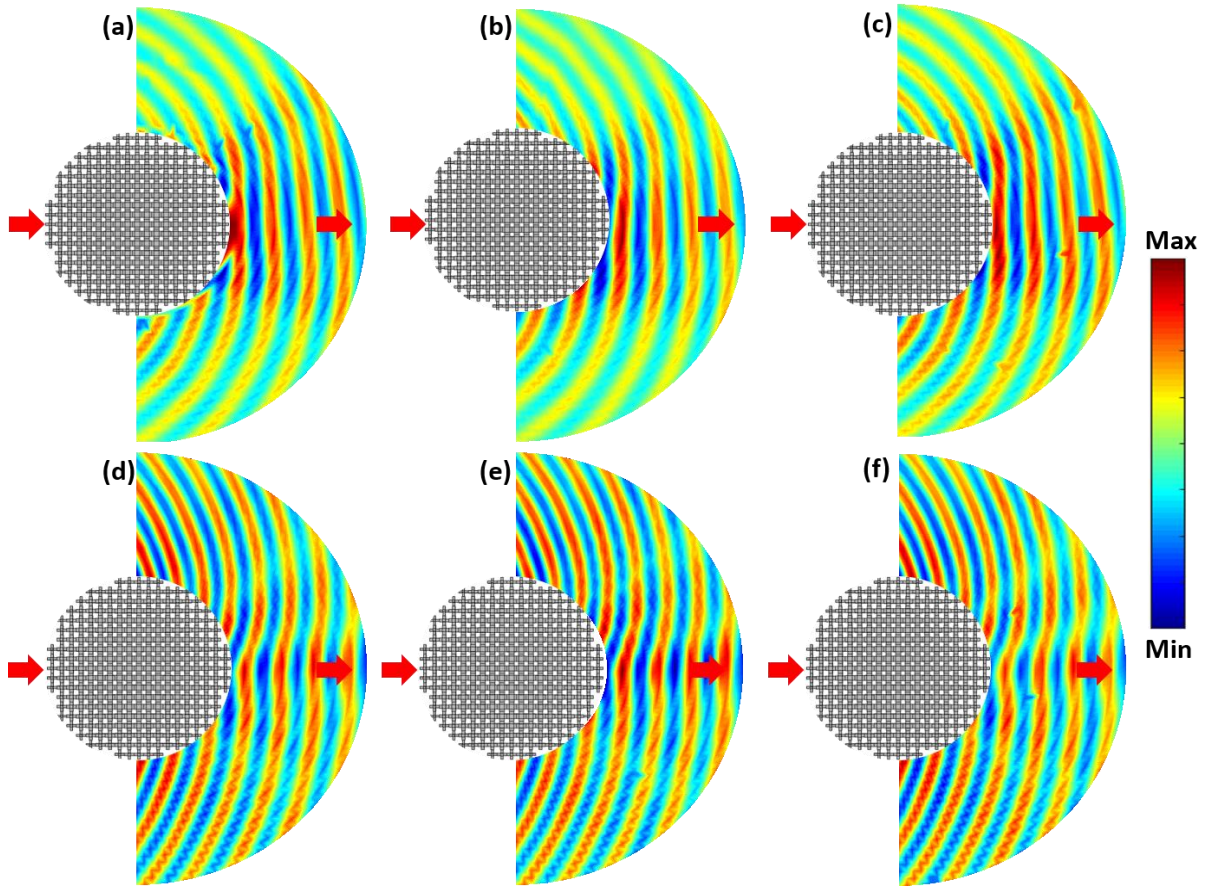

**Figure 8:** Experimental results for acoustic collimation and focusing using 3D lens. (a)-(c) Acoustic collimation using 3D RAALL at the frequency $f$ = 40 kHz with measurements performed at $z$ = -2 mm, 0 mm, and 2 mm locations. (d)-(f) Acoustic focusing using 3D RAALL at the frequency $f$ = 40 kHz with measurements performed at $z$ = -2 mm, 0 mm, and 2 mm locations.

## 5. Conclusions

We theoretically, numerically and experimentally design and test both 2D and 3D reduced aberration acoustics Luneburg lens (RAALL) for acoustic collimation and acoustic focusing. The Luneburg lens is based on the gradually change of the refractive index, which can smoothly change the wave propagation direction. We modified the governing equations of traditional acoustic Luneburg lens in order to achieve broadband and omnidirectional characteristics of the lens. Both 2D and 3D RAALLs are fabricated using 3D printer and tested experimentally. The experimental acoustic collimation and focusing performances are consistent with numerical simulations. The RAALL could benefit many potential applications for ultrasonic imaging, diagnosis, treatment, sonar system and energy harvesting.

**Acknowledgements**

**Conflict of Interest**

The authors declare no conflict of interest.